# SMOOTHING AND MODELING OF VIDEO TRANSMISSION RATES OVER A QOS NETWORK WITH LIMITED BANDWIDTH CONNECTIONS

Khaled Shuaib, Farag Sallabi and Liren Zhang

Faculty of Information Technology, United Arab Emirates University

k.shuaib@uaeu.ac.ae, f.sallabi@uaeu.ac.ae, lzhang@uaeu.ac.ae

### ABSTRACT

*Transmission of video over a limited bandwidth network is challenging due to the natural variability of video, and link characteristics. Video smoothing techniques can be used to facilitate more effective transmission and to preserve better quality. In this paper we develop a semi-optimal video smoothing approach to manage the transmission of MPEG-4 and H.264 video while mapping it to be more suitable for a QoS based network. The proposed technique utilizes a smoothing buffer with pre-defined thresholds to smooth the transmission rates while assuming minimal information about the video to be transmitted. The results obtained showed a significant improvements in transmission rate variability while guaranteeing no buffer overflows or underflows. In addition, a queuing model is developed for the used smoothing algorithm for H.264 video streams with optimized encoding and packetization, utilizing the available H.264 macroblock ordering option.*

## 1. INTRODUCTION

Subscriptions to broadband services are expected to reach 1.8 billion by 2012. High data rates, improved performance and user experience will be the trend driving the development of fixed and mobile broadband technologies for applications such as IP interactive TV, video, gaming, e-health and others [1, 2]. These applications have different QoS requirements to satisfy the intended users' expectations and therefore QoS-enabled networks play a crucial role in the successful deployment of such applications. The main objective of a QoS-enabled network infrastructure is to ensure that the users get the desired experience they expect based on their service level agreement with the operator that owns and manages the network. On the other hand, from the operator point of view, applying QoS implies that it can optimize usage of limited network resources while satisfying customers. Among the above mentioned applications, video is considered to be the most challenging.

When a video stream is encoded as variable bit rate (VBR), bit allocation and distribution is varied depending on the complexity and motion of each scene. This is done to obtain an optimal video quality while not consuming more resources than needed. The video variability is very hard to measure and depends on the chosen encoding parameters of the video clip, mainly the mean encoding bit rate (CER) and the peak encoding bit rate (PER). The greater is the difference between these two parameters, the greater is the assumed variability in the video stream which results in great frame size variability. To alleviate this variability for the transmission of video over limited bandwidth network connections, and for better provisioning of network resources certain measures are needed. To achieve this, traffic classifiers and conditioners, also called traffic shapers or smoothers have been proposed by researches before [3, 4, 5]. Where traffic classifiers can be used to prioritize traffic types, the main concept applied by traffic shapers is to use one or more shaping buffers managed by particular algorithms at different places between the sender and the receiver to control and adapt the rate at which the traffic is being sent, and therefore help comply with the conditions of the network connections/channels used or reserved. Figure 1 shows a network diagram of how video can be transported over a wired/wireless network where the smoothing of video can be done at the source or just before the air interface i.e. NodeB in the case of a WiMAX or LTE wireless network.

The rest of the paper is organized as follow. In section two related work is presented and in section three the proposed video smoothing technique is discussed. Section four presents a system model and section five shows and discusses the obtained performance results. In section six a queuing model for H.264 video is presented and the paper is then concluded in section seven.

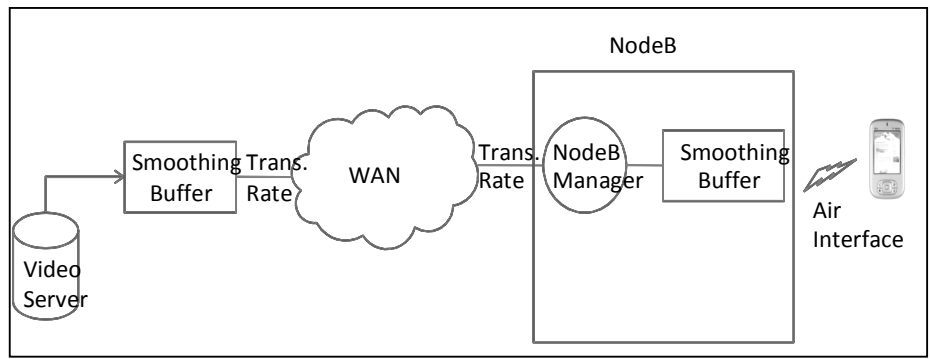

Figure 1. Video Transmission over a network

## 2. RELATED WORK

Two main extreme techniques of video smoothing have been mainly used in the literature: basic smoothing and optimal smoothing [6, 7]. In basic smoothing, video is transmitted at the average rate of N none overlapping successive frames. In this technique, the larger is N the less variability there is, but the larger is the smoothing delay. In optimal smoothing which has a greater complexity, the transmitting bit rate is minimized while guaranteeing an upper bound on delay and no over/under flows of the decoder buffer. This is achieved by using piecewise constant bit rate segments which are as long as possible. Optimal smoothing is only suitable for pre-encoded video since it computes the transmission rate schedule off line. In both schemes the transmission rate is changed for every frame which is not appropriate for all kind of communication networks. For transporting pre-stored video over unmanaged IP networks, the authors in [8] proposed a rate control technique based on modern control theory to optimize the user-perceived video which was measured using an evaluation function. The technique was mainly based on controlling both the sending rate and video rate to achieve optimized results. In another paper [9], the authors focused on the performance evaluation of MPEG-4 video transmission based on their own proposed single-rate multicast transport protocol for multimedia applications. The authors primarily focused on a detailed evaluation which was based on both network-centric and video quality metrics and concluded that moderate and stable transmission rates of video with minimum losses provided a better service to the end user in terms of video quality. To smooth out the video play-out for better quality viewing when transmitted over a wireless channel and to improve the overall system performance, the authors in [10] proposed a new adaptive media play-out (AMP) system based on a packet-delay prediction algorithm which makes decisions based on the delay interdependency of the adjacent packets. Most recently, the authors in [11] presented a theoretical window based model to achieve a tradeoff between buffer occupancy and picture quality. This was gracefully achieved by regulating the size of the window composed of several adjacent frames.

Although scheduling, admission control and proper resource allocation are not within the scope of this paper, however it plays a vital role in QoS networks, especially emerging wireless networks as was discussed by other researchers. Several researchers have worked on call admission, scheduling and reservation protocols for the various types of networks. For example, the authors in [12] presented an optimization-based formulation for scheduling and resource allocation in the uplink OFDM access network which can also be applied to an LTE network. In another paper the authors in [13] were able to show how proper scheduling and resource reservation can be used for pre-encoded video stream over wireless downlink packet networks while incorporating information obtained from the video decoder for the optimization of resource allocation and the preservation of video quality. In [14] the authors presented an adaptive and fair priority scheduler for video streaming over wireless links where they considered the characteristics and strict requirements of multimedia applications being transmitted. Furthermore, the authors in [15] proposed and evaluated two scheduling algorithms tailored toward OFDM based broadband wireless systems. In [16] the authors combined admission control and scheduling based on QoS differentiation of a mixed traffic scenario in an LTE uplink. For multimedia traffic, the authors in [17] proposed a new Data Link Layer protocol which grants a source permission to transmit over a wireless channel based on a priority scheme that takes into account the time-to-live parameter of all the transactions, selectable priorities assigned to all the sources and the relevant channel condition.

## 3. PROPOSED VIDEO SMOOTHING

In this paper a semi-optimal smoothing technique that can be used for both pre-encoded and real-time video is proposed. The technique uses two buffer thresholds to manage the transmission rate (R) while guaranteeing no buffer over/under flow and a desired start-up delay. This model can be interpreted using a three state Markov chain process with transition being possible only between adjacent states (R1, CER_t, R2) as shown in Figure 2. Where R1 is a transmission bit rate less than the mean transmission bit rate (CER_t) and R2 is a transmission bit rate greater than CER_t, but less or equal to PER. One of the proposed scheme objectives is to minimize the transmission at R2 and to transmit at CER_t or R1 whenever possible. This is to a align R= CER_t with the use of a constant bit rate (CBR) when going over an network where data rates can be guaranteed up to a pre-chosen rate = CBR. When R is anticipated to be greater or less than CBR, a new rate matching R will be negotiated between the edge network node and the client to guarantee conformance with the agreed upon service level agreement. This could be based on a traffic contract that can be agreed upon between the ISP and the client to allow the transmission at a rate above the CBR when needed with proper billing to account for the client receiving information below or above R. The contract can have provisioning for credit being given when R < CBR and dept when R > CBR. Once the transmission is finished or stopped, a billing statement based on the agreed upon traffic contract can be generated to account for any additional dept or credit. In the proposed smoothing scheme, CER_t can be chosen initially based on the CER value and can be expressed as: CER_t = CER (1+$\alpha$) bps, where $\alpha$ is a video variability factor greater or equal to zero and can be chosen initially by the user for real-time transmissions or through video analysis for pre-encoded streams. From our simulation of pre-encoded video a value between 0.01 and 0.03 was needed to compensate for the difference between the gained credit and owed dept by the end of the transmission when smoothing is used. Another way to choose $\alpha$ would be based on the maximum allowed smoothing buffer delay.

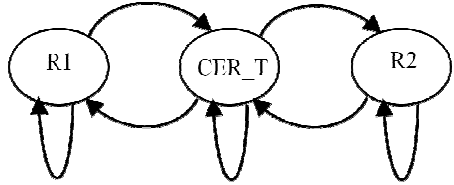

Figure 2. Transmission rate states

## 4. SYSTEM MODEL

The performance of the proposed video smoothing approach was evaluated using a simulation program written in Java. The simulation program is based on a client/server paradigm. At the server, the video frames are generated every 33 ms, i.e. every frame period (Tf), and saved at a FIFO synchronized queue serving as a smoothing buffer. Data is then read from the queue based on R utilizing an RTP frame length of 30 ms for transmission to the client which acted as a sink. Several MPEG-4 and H.264 video traces were used in the simulation. These video traces and their statistics were obtained from [18, 19, 20] and summarized in Table 2. The video traces were chosen to represent various video types (sport, movie, news). Corresponding MPEG-4 and H.264 traces used were best chosen for similar mean frame peak signal to noise ratio (PSNR) so that a better and fair comparison can be made. All traces were encoded as 30 frames per second VBR CIF 352x288 with a group of picture (GOP) defined as G16B3 i.e. IBBBPBBBPBBBPBBBP [18, 19, 20].

Table 2. Video traces used in this work

| Video Trace | Mean Frame Bit Rate (bps) | Peak Frame Bit Rate (bps) | Number of Frames | Mean Frame PSNR (db) |
|---|---|---|---|---|
| H.264-Tokyo Olympics | 144267.6 | 4182960 | 133125 | 35.557 |
| MPEG4- Tokyo Olympics | 278896.9 | 3358080 | 133125 | 34.829 |
| H.264- Silence of the Lambs | 68898.9 | 3168720 | 53997 | 37.598 |
| MPEG4- Silence of the Lambs | 194176.9 | 2592480 | 53953 | 37.116 |
| H.264- NBC 12 News | 197427.6 | 3393840 | 49521 | 33.131 |
| MPEG4- NBC 12 News | 420051.4 | 4248240 | 49521 | 33.375 |

In the proposed scheme minimal information about the video traces was assumed to be known for applying smoothing, making it suitable for both pre-encoded and real-time video. The steps of the smoothing scheme algorithm are outlined in Figure 3. $A_1$ in the algorithm is the first buffer occupancy threshold chosen to avoid any buffer underflow as R is chosen to guarantee that the maximum amount to be transmitted is no more than the current content of the buffer (B). $A_2$ is the second buffer occupancy threshold chosen to keep the startup delay down to a desired value, to maximize the transmission rate at the CER_t, to minimize the transmission at the PER, to avoid any buffer overflow and to keep the maximum buffering delay (D in seconds) below a certain desired limit. D in this case can be expressed by:

D= [($A_2$/CER_t * 8) + Tf] or D= Tf * [(PER/CER_t) + 1]

By limiting the maximum smoothing buffer delay to D, one can calculate the expected CER_t and therefore the needed value of α based on CER_t = CER (1+α) as was indicated earlier.

The proposed algorithm is considered to be semi-optimal for two reasons: 1) the maximum buffering delay is not linked to the playback time of the video frames as more smoothing can be done at the receiving end to overcome that. 2) the instantaneous amount of net credit/dept might not be zero i.e. the number of transmitted bytes at any instant of time is not conformant to the agreed upon average transmission rate. However, this should have been agreed upon before the start of any transmission and a final billing statement can be produced once the transmission is over to make for any needed adjustments. On the other hand, and for a risk of introducing additional delay i.e. larger maximum buffer fullness, step 3c in Figure 3 can be modified to guarantee that R will not be above the agreed upon CER_t unless there is available net credit. In this case R will be tied to the number of bytes available as credit and step 3c will be modified as in Figure 4. In the next section, results are obtained to investigate the performance of the proposed algorithm.

1. Choose two thresholds ($A_1$ and $A_2$) where
   $A_1$ = (CER_t /8)* Tf;
   $A_2$ = (PER /8) * Tf;
2. Pre-fill the buffer with video data until $A_1$ before starting to transmit any data over the network
3. Choose a transmission rate, R, based on the buffer fullness (B) in bytes as follow:
   a) If {B < $A_1$}, then
      R = (B / Tf)*8 bps;
      Credit = Credit + [(CER_t – R)/8] bytes;
   b) Else If {$A_1$ <= B <= $A_2$} then
      R = CER_t bps;
   c) Else (i.e. B > $A_2$), then
      R = Max. [CER_t, (Min. (PER, ((B-$A_2$)/ Tf) * 8) bps];
      Dept = Dept + [(R – CER_t)/8] bytes;
4. Read a video frame into the buffer every Tf
5. Transmit an RTP frame every RTP frame period
6. Go back to 3 and repeat until there are no video frames left to transmit.
7. When done, generate a final billing statement. This will be based on the agreed upon mean transmission rate and any difference between the gained credit and owed dept from step 3.

Figure 3. The proposed smoothing algorithm

```
If (Credit – Dept) > 0 {
R = Max. [CER_t, (Min. (PER, ((Credit-Dept)/Tf)*8) bps];
Dept = Dept + [(R – CER_t)/8] bytes;
Credit = Credit – [(R – CER_t)/8] bytes;
}
Else {R = CER_t};
```

Figure 4. R based on the availability of credit

## 5. RESULTS AND ANALYSIS

To look at the performance of the proposed algorithm, several experiments were conducted. Table 3 shows the general performance results when CER_t was set to be CER. As can be seen from Table 3, when the smoothing technique is applied the value of R is roughly 50% of the time around the CER_t except for the H.264 NBC 12 News which is 67%. Adding up the percentages of time when R is either at R1 or CER reflects the percentage of time where transmission is <= CBR in a limited bandwidth network. The results also reveal that a CER_t greater than the CER is needed to minimize the (Dept – Credit) value for optimal overall transmission based on a traffic contract.

In Table 4, we looked at the effect of changing the value of $A_1$. As can be seen from the results of Table 4, increasing the value of $A_1$ had no significant impact on the performance except for an additional start up delay since the smoothing buffer need to be pre-filled to the first buffer threshold before transmission. In Table 5, we looked at the effect of choosing a CER_t greater than CER. This is to show how choosing a value of $\alpha$ will affect the performance. The value of $\alpha$ here was chosen assuming that a pre-analysis phase was done which resulted in the calculation of that value. The value used was based on the ratio (CER_t Obtained / CER) from Table 3. Where CER_t Obtained refers to the average transmission rate obtained from the analysis phase. The main objective is to reduce the value of the (Dept – Credit) bytes per second as can be seen from Table 5 to better conform to the agreed upon traffic contract.

Table 3. Performance results for transmitting video traces

| Video Trace | (Dept – Credit) bytes per second | Percentage at R1, CER, R2 | Maximum B in bytes | CER_t Obtained /CER |
|---|---|---|---|---|
| H.264-Tokyo Olympics | 198.2 | 34.4, 52.6, 13.0 | 34514 | 1.0116 |
| MPEG4- Tokyo Olympics | 406.9 | 29.0, 48.9, 22.1 | 28142 | 1.0116 |
| H.264- Silence of the Lambs | 124.7 | 34.2, 52.7, 13.1 | 24986 | 1.0149 |
| MPEG4- Silence of the Lambs | 166.7 | 29.2, 50.9, 19.9 | 20910 | 1.010 |
| H.264- NBC 12 News | 285.8 | 23.5, 67.0, 9.5 | 26461 | 1.0118 |
| MPEG4- NBC 12 News | 524.3 | 28.3, 53.8, 17.9 | 32784 | 1.010 |

Table 4. H.264- Silence of the Lambs transmitted at R = CER for different pre-fill buffer values

| Pre-Fill Value in bytes | (Dept – Credit) bytes per second | Percentage at R1, CER, R2 | Maximum B in bytes | Obtained CER_t |
|---|---|---|---|---|
| $A_1$ | 124.7 | 34.3, 57.8, 7.9 | 24986 | 1.0149*CER |
| 50 $A_1$ | 121.4 | 34.3, 57.9, 7.8 | 24998 | 1.0145*CER |

Table 5. Using obtained CER_t instead of CER

| R | (Dept – Credit) bytes per sec | Percentage at R1, CER, R2 | Video Trace |
|---|---|---|---|
| CER | 124.7 | 34.3, 57.8, 7.9 | H.264- Silence of the Lambs |
| 1.0149 * CER | - (35.4) | 35.3, 57.0, 7.7 | |
| CER | 166.7 | 29.2, 50.9, 19.9 | MPEG4- Silence of the Lambs |
| 1.010* CER | 59.5 | 29.4, 51.1, 19.6 | |

To look at the effect of smoothing on the rate variability we used the variability definition from (G. Van et.al 2008) given by:

V = Standard Deviation of Transmission Rates / Mean Transmission Rate.

Figure 5 shows the results on variability for six video traces when transmitted using the proposed smoothing technique and without any smoothing applied i.e. each video frame is being transmitted using a transmission rate calculated based on the frame size. As can be seen from Figure 5, improvement in rate variability is above 20% for all traces with more than 46% for the H.264 NBC 12 News clip. This is mainly due to the fact that R is at R1 or CER the majority of the time. This can also be captured from Figure 6 which shows a sample of 100 transmission rates for the H.264 Silence of the Lamb video trace.

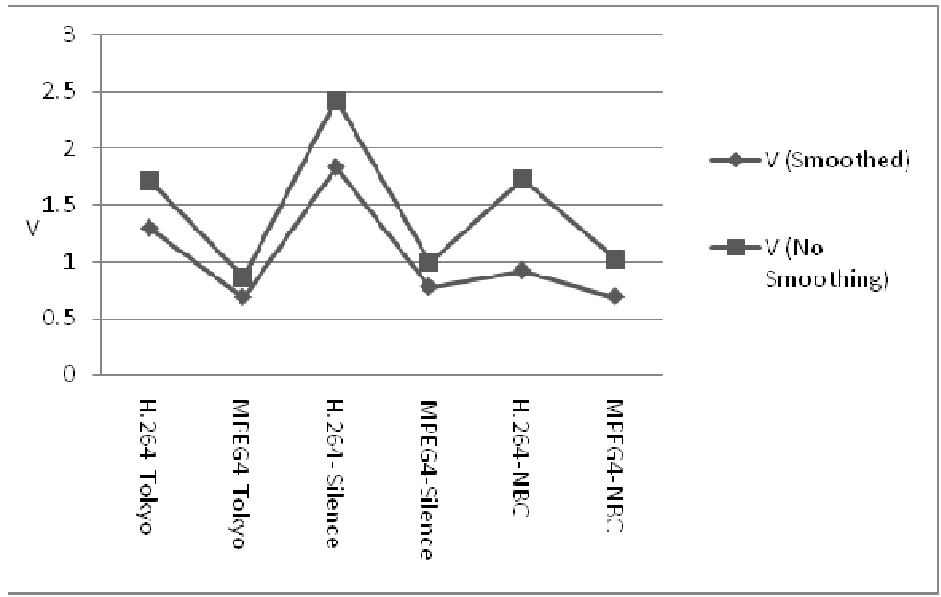

Figure 5. Effect of smoothing on the variability of transmission rate

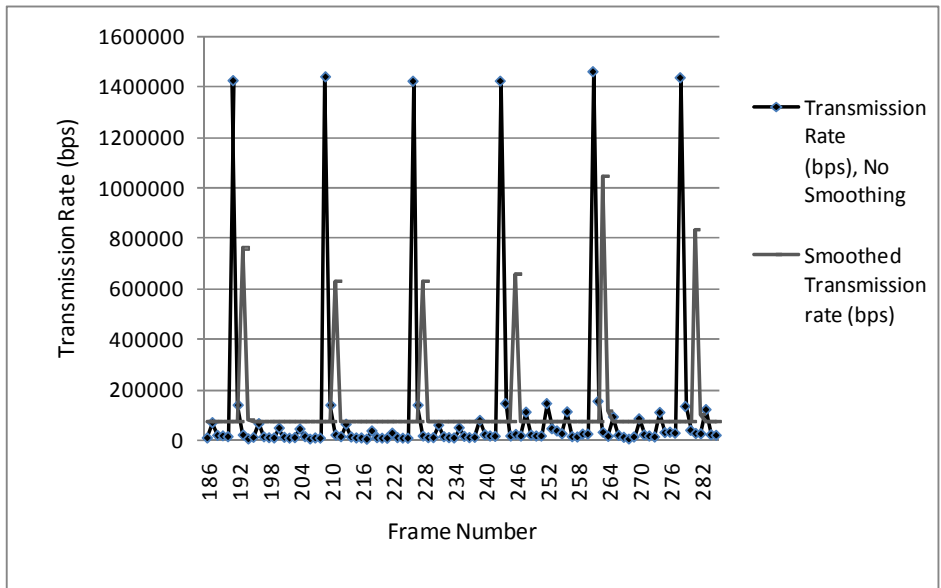

Figure 6. Effect of smoothing on transmission rates

154

Finally, we looked at the performance of the algorithm with step 3c in Figure 3 being replaced with the one in Figure 4. In this case, R was tied to the availability of credit when the buffer content was above $A_2$. This case was run for the H.264 Silence of the Lamb clip and the results showed V = 2.4 and 172 buffer overflow instances indicating a poor performance. In the next subsection we look at the performance of the smoothing scheme when feedback on bandwidth is provided from the network for the connection over which the video is being transported.

**5.1 Smoothing with Network Connection Feedback**

The average/maximum combination of available bandwidth of a network video connection over the period of an RTP frame, 30 ms, as can be allocated by the scheduler at the edge node, i.e. NodeB in case of a cellular network such as LTE, is fed back to the smoothing buffer and used to regulate the transmission rate. In this case R in the smoothing algorithm will not just depend on the fullness of the smoothing buffer, but will also depend on the available connection bandwidth parameters. We assume that the average connection bandwidth (Rc) and the maximum available connection bandwidth (Rmax) do not change within the length of an RTP frame i.e. 30 ms. For this mode, step 3 in the algorithm of Figure 3 is modified as shown in Figure 7. As an example, we have used data generated from the proposed LTE wireless channel model in [21] for a system bandwidth of 10 MHz. We assume that the scheduler at NodeB is to allocate an average wireless channel bandwidth based on the average bit rate of Physical Resource Blocks (PRBs) over a number of LTE physical layer transmission time intervals (TTIs), within an LTE frame of 10 ms. Based on this proposed model the generated average bit rate per PRB is shown in Figure 8. Rc and Rmax in this case are chosen to be as close as possible to the encoding parameters of the video clip being transmitted. For example, when transmitting the H.264 Silence of the Lambs clip which has a CER around 70 Kbps, Rc was based on the allocation one PRB every 4 TTIs within an LTE frame and Rmax was based on the allocation of 9 PRBs per TTI. Based on this, the average PRB bit rate generated, Rc, will be around 80 Kbps and the average maximum bit rate that can be allocated, Rmax, will be around 2.9 Mbps. The transmission rate variability results for the Silence of the Lambs H.264 are shown in Figure 9. As can be seen, the transmission rate variability was further reduced with channel feedback from 1.8 to 1.4 and no buffer over/under flows occurred.

> Choose a transmission rate, R, based on the buffer fullness (B) in bytes and the average channel bandwidth as follow:
> a) If $\{B < A_1\}$, then
>    R = Min ((B / Tf)*8 bps), Rc);
>    Credit = Credit + [(CER_t – R)/8] bytes;
> b) Else If $\{A_1 <= B <= A_2\}$ then
>    R = Min (CER_t, Rc) bps;
> c) Else (i.e. $B > A_2$), then
>    R = Min {Max [CER_t, (Min. (PER, ((B- $A_2$)/ Tf) * 8) ], Rmax} bps;
>    Debt = Debt + [(R – CER_t)/8] bytes;

Figure 7. Step 3 of the smoothing algorithm with feedback

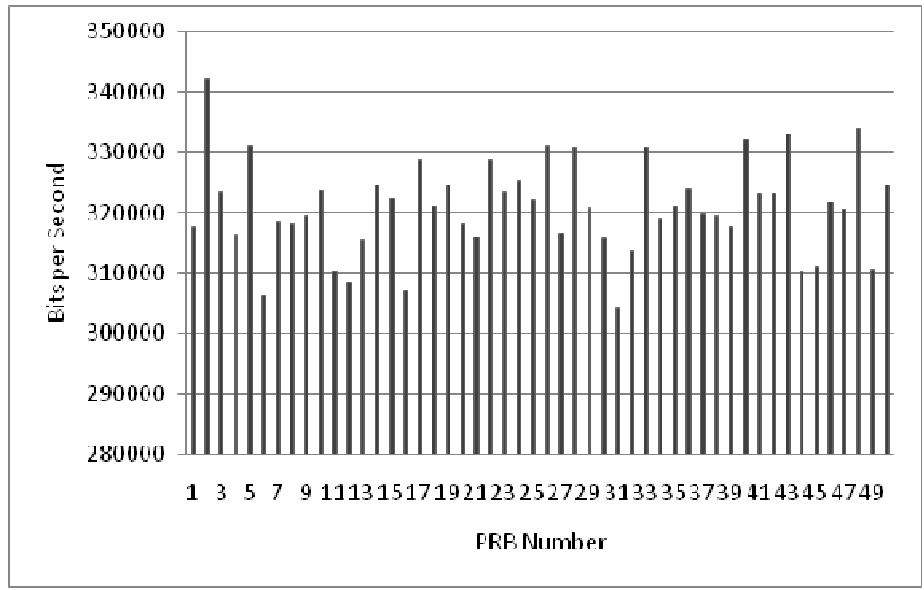

Figure 8. Average PRB bit rate per TTI

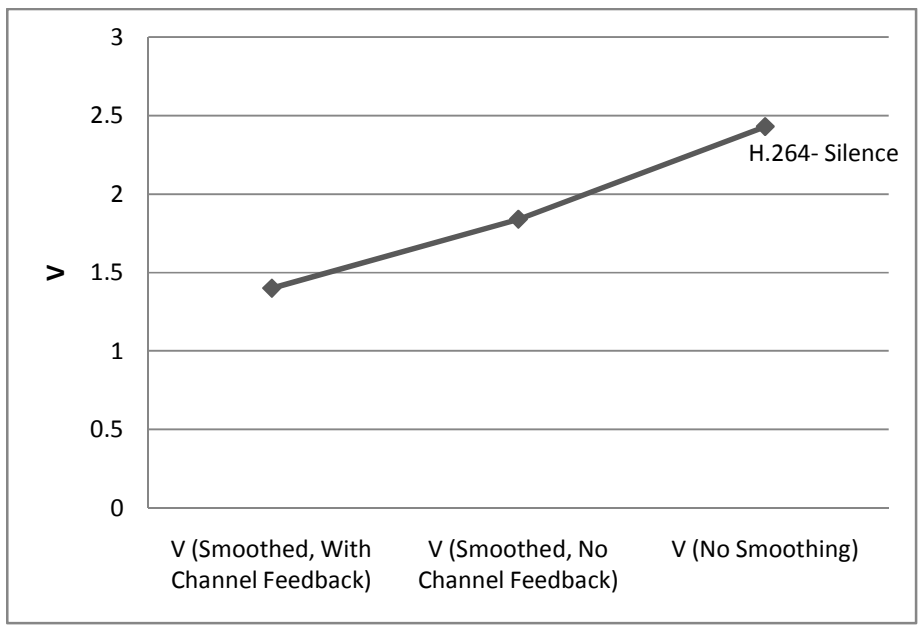

Figure 9. Variability results with and without channel feedback

## 6. MODELING OF SMOOTHING H.264 VIDEO

The H.264 video traces are considered in this paper to be encoded by taking the advantage of the flexible macroblock ordering (FMO) option [22], which allows macroblocks relevant to the same object within each video frame to be grouped together [23]. This feature is achieved by adding a new layer, called macroblock groups, in video frames. With the FMO option, an H.264 video encoder is not any more restricted to raster scan order. Instead, the encoder first assigns macroblocks located in a video frame into macroblock groups and then it packetizes each macroblock group independently into a single packet. Hence, after encoding and packetizing, the number of packets contained in a video frame is a random variable, which

156

depends on the number of macroblock groups contained in each video frame. H.264 video traces are assumed to be generated and transmitted at a rate of 30 frames per second. In the following analysis, we assume that each video frame consists of a maximum of N independent macroblock groups, where each macroblock group is modeled as an ON-OFF process that generates a packet when it is in the ON state. As shown in Figure 10, the transition rate from ON state to OFF state is ρ and the transition rate from OFF state to ON state is β. When N of such ON-OFF packet streams are multiplexed, the resultant packet stream can be represented by an (N+1) state Markov modulated process as shown in Figure 11, where the state i represents that i ( i = 0, 1, 2, …, N) packets in a video frame. The transition rate from the state i to the state (i-1) is iβ and the transition rate from the state i to the state (i+1) is (N-i)ρ

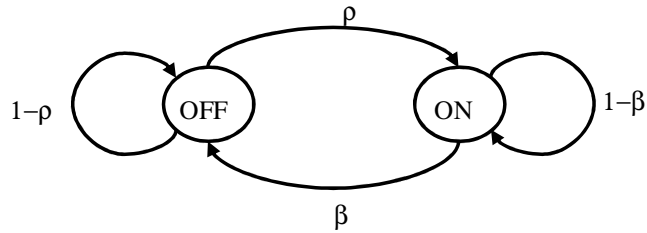

Figure 10. On-Off Traffic Model

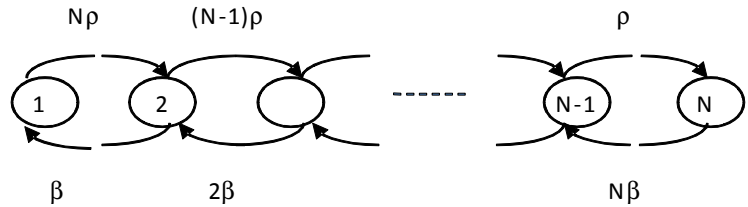

Figure 11. Multiplexing of ON-OFF Sources

## 6.1 Performance Analysis of the Proposed Model

As shown in Figure 12, the queuing model of a smoothing buffer has a buffering capacity of K packets with thresholds $A_1$ and $A_2$, which are used for service rate control. Let x be the number of packets in the queuing system. When $0 \leq x \leq A_1$, the service rate is $CER_t(1-\alpha)$ packets per second. Likewise, when $A_1 \leq x < A_2$, and $A_2 \leq x \leq K$ the service rates are $CER_t$ and $CER_t(1+\alpha)$ packets per second, respectively, where $(0 \leq \alpha \leq 1)$ is a video variability factor and can be chosen initially by the user for real-time transmissions or through video analysis for pre-encoded streams.

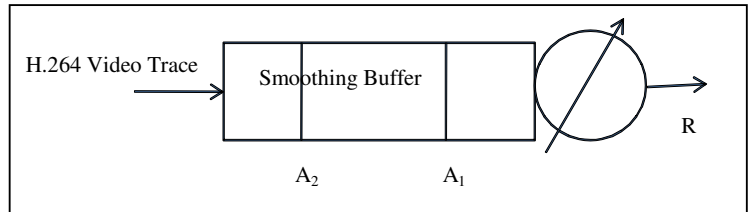

Figure 12. Management of transmission rates

The analysis here is based on a fluid flow model [24] since packets arrive at the input of the queuing system in bursts given by $\binom{N}{i} p_{ON}^i (1-p_{ON})^{N-i}$ where i represents the number of ON-OFF processes among the N processes in the ON state and $p_{ON}$ represents the probability that an ON-OFF process is in the ON state. Hence, the classic solution based on a Markov chain model is not applicable. The following analysis is based on fluid flow technique given in [24] with the assumption that N>>1 and the output transmission link is divided into equal time slots and each slot is equivalent to the maximum transmission time of a packet such that the transmission time of RTP/IP packets is assumed to be uniformly distributed. This assumption is reasonable, because when N>>1, the bit rate allocated to each packet when compared to the large capacity of the broadband transmission link is negligible. Therefore, variation of the transmission time from one packet to another is negligible. Considering that the input stream consists of i ON-OFF processes that are in the ON state at time t, we define $F_i(t,x)$ ( $0 \le i \le N, A_m \le x < A_{m+1}, m \in \{0,1,2\}$ ), as the cumulative probability distribution for the packets in the queue at time t, where i out of N ON-OFF processes are in the ON state. In fact, $F_i(t,x)$ represents the probability that the queuing buffer occupancy is less than or equal to x for ( $0 \le x < A_1$ ), ($A_1 \le x < A_2$ ) and ( $A_2 \le x < K$ ) while i processes are in the ON state at time t. According to Figure 12, $F_i(t, x)$ can be calculated by setting up a generating equation of $F_i(t+\Delta t, x)$ which is the probability at an incremental time of t+$\Delta$t. Then, we have

$$F_i(t+\Delta t, x) = [N-(i-1)]\rho \Delta t F_{i-1}(t,x) + (i+1)\beta \Delta t F_{i+1}(t,x)$$
$$+ \{1-[(N-i)\rho + i\beta]\Delta t\} F_i[t, x-(i\lambda - C_x)\Delta t] + o(\Delta t) \qquad (1)$$

$$C_x = \begin{cases} CER_t(1-\alpha) & 0 \le x \le A_1 \\ CER_t & A_1 \le x \le A_2 \\ CER_t(1+\alpha) & A_2 \le x \le K \end{cases}$$

with the term $x - (i\lambda - C_x)\Delta t$ being the buffer occupancy. On the right side of equation (1), the first term is the probability of transition from the state (i-1) to the state i at time t+$\Delta$t, the second term is the probability of transition from the state (i+1) to the state i and the third term is the probability that the system is at state I and is not changing at time t. The term o($\Delta$t) represents all higher order terms which go to zero much rapidly than $\Delta$t when $\Delta$t tends to zero. Hence, the effects of o($\Delta$t) is negligible when $\Delta$t is small enough. In equation (1), we also assume that $F_{-1}(t,x)$ and $F_{N+1}(t,x)$ are set to zero.

Now we expand $F_i(t+\Delta t,x)$ and $F_i(t,x-\Delta x)$ for $\Delta x = (i\lambda - C_x)\Delta t$ in their respective Taylor series with the assumption that the appropriate continuity conditions are satisfied. Let $\Delta$t goes to zero, then equation (1) represents the following differential function:

$$(i\lambda - C_x) d(F_i(x))/dx = [N-(i-1)]\rho F_{i-1}(x)$$
$$+ (i+1)\beta F_{i+1}(x) - [(N-i)\rho + i\beta] F_i(x) \qquad (2)$$
$$0 \le i \le N, \ F_{-1}(x) = 0, \ F_{N+1} = 0$$

Define $F(x) \equiv [F_0(x), F_1(x), \ldots F_N(x)]^T$, then equation (2) can be expressed in the following compact matrix form:

$$D \, dF(x)/dx = MF(x) \qquad (3)$$

Where D is a tri-diagonal matrix defined as
$$D = diag[-C_x, \lambda - C_x, \ldots, N\lambda - C_x], \text{ M is the } (N+1) \times (N+1)$$

158

Assuming $(i\lambda - C_x)$ is not equal to zero for any i, $(0 \le i \le N)$, the general solution of equation (3) can be given by

$$F(x) = \sum_{j=0}^{N} a_j V_j e^{z_j x}, \qquad (4)$$

Where the elements in the vector $z [z_0, z_1, \ldots z_N]$ are the Eigen values and $V_j$ is the Eigen vector of the matrix $(D)^{-1}M$. In equation (4), the coefficients $\{a_j\}$ can be obtained from the boundary conditions by defining $E_D = \{i | i\lambda < C_x\}$ and $E_u = \{i | i\lambda > C_x\}$. By doing that, the boundary conditions in equation (4) can then be obtained as below where $P_i = \binom{N}{i} P_{on}^i (1 - P_{on})^{N-i}$ is the probability that i sources are in the ON-state and $P_{on} = \rho/(\rho + \beta)$ is the probability that a source is in the ON-state. Hence, the steady-state distributions under the boundary conditions of equation (5) can be used to calculate the throughput for the traffic with different threshold values for buffer occupancy control as in equation (6).

$$\begin{aligned} F_i(0) &= 0 & &\text{if } i \in E_u, \\ F_i(A_m) &= P_i & &\text{if } i \in E_D, \quad (5) \\ F_i(A_m) &= F_i(A_m) & &\text{if } i \in E_u \cup E_D, \end{aligned}$$

$$T = \sum_{i=0}^{N} F_i(A_m) i\lambda - \sum_{i \in E_u \cap E_D} \left[ (i\lambda - C_x)(F_i(A_{m-1}) - F_i(A_m)) \right] \qquad (6)$$

With the traffic arrival rate given by $A = \sum_{i=0}^{N} i\lambda P_i$, then the packet loss probability due to buffer overflow will be given by $P_L = 1 - T/A$. Likewise, the probability that the output link speed at $CER_t(1-\alpha)$, $CER_t$ and $CER_t(1+\alpha)$ can be calculated using equation (4) based on the boundary conditions presented in equation (5). That is

$$P_r(0 \le x \le A_1) = \sum_{x=0}^{A_1} F(x) = \sum_{x=0}^{A_1} \sum_{j=0}^{N} a_j V_j e^{z_j x} \qquad (7)$$

$$P_r(A_1 \le x \le A_2) = \sum_{x=A_1}^{A_2} F(x) = \sum_{x=A_1}^{A_2} \sum_{j=0}^{N} a_j V_j e^{z_j x} \qquad (8)$$

$$P_r(A_2 \le x \le K) = \sum_{x=A_2}^{K} F(x) = \sum_{x=A_2}^{K} \sum_{j=0}^{N} a_j V_j e^{z_j x} \qquad (9)$$

In addition, the probability distribution for a buffer occupancy less than $A_m, (m=0,1,2)$ is given by

$$P_r\{delay = A_m / C_x\} = 1/T \left(1 - \sum_{i=0}^{N} F_i(A_1)\right) C_x \qquad (10)$$

## 7. CONCLUSIONS

In this paper a video smoothing technique is proposed for the transmission of video over a limited bandwidth network. An evaluation of the proposed scheme under different scenarios was conducted and the results obtained showed good improvements in transmission rate variability while having no losses due to smoothing. In particular, a detailed example of the transmission of video over an LTE wireless network was considered to provide better insight into how the variability of the wireless channel can be taken into consideration when video smoothing is performed by the proposed technique. Several video traces of MPEG-4 and H.264 with different characteristics were used in the evaluation process and the results reflected how the H.264 video exhibited more variability than MPEG-4 at similar encoding rates. In addition, a queuing model of the smoothing scheme is constructed for H.264 video streams which can be generalized to multiple input sources of video and can be used to model video smoothing for applications such as IP TV and video distribution over a heterogeneous network.

**Authors**

Khaled Shuaib received his PhD in Electrical Engineering from the Graduate Center of the City University of New York, 1999, and his MS and BS in Electrical Engineering from the City College of New York, 1993 and 1991 respectively. Since September 2002, Khaled has been with the Faculty of Information Technology, at the UAEU where he is currently an Associate Professor and Associate Dean for Student Affairs. His research interests are in the area of network design and performance, networking protocols, sensor networks, multimedia transmission over 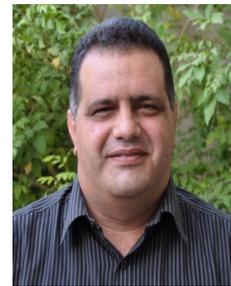 wireless networks, network security and networking for the Smart Grid. Khaled is a Senior Member of the IEEE and serves on the technical program committees of international IEEE conferences such as GLOBECOM, ISCC and ICC. He is also on the editorial board of the International Journal of Informational Sciences and Computer Engineering, and the International Journal of Computer Science and Network Security.  Prior to joining the UAEU, Khaled had several years of industrial experience in the US working as a Senior Member of Technical staff at GTE Labs (Currently Verizon Technology Center) (1997-1999), and as a Principle Performance Engineer for Lucent Technologies (1999-2002).

Farag Sallabi is an Associate Professor of Networking at the Faculty of Information Technology, United Arab Emirates University. Dr. Sallabi joined the UAE University in September 2002. He is involved in several research projects that investigate the Quality of Service provisioning in wired and wireless networks, Routing in Ad Hoc Wireless networks and design and implementation of Real-time embedded systems. Before joining the UAEU, Dr. Sallabi worked at Sigpro Wireless in Canada, a company that develops and implements software and hardware for 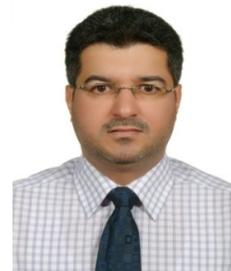 handheld devices. Dr. Sallabi received his Ph.D. in Electrical and Computer Engineering from University of Ottawa, Canada in 2002. During his PhD studies, he developed and implemented protocols for providing Quality of Service for multimedia application in the Internet. Dr.

Sallabi received his B.Sc. and M.Sc. Degrees in Electrical and Electronic Engineering from Garyounis University, Benghazi, Libya in 1987 and1995 respectively.

Liren Zhang received his M.Eng. (1988) from the University of South Australia and Ph.D (1993) from the University of Adelaide, Australia, all in electronics and computer systems engineering. He joined the Faculty of Information Technology, UAEU as Professor in Networking in 2009. Dr Zhang was a Senior Lecturer in Monash University, Australia (1990 – 1995), an Associate Professor in Nanyang Technological University, Singapore (1995 – 2007) and a Research Professor of Systems Performance Modeling in University of South 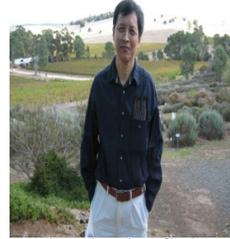 Australia. Dr. Zhang has vast experience as an engineer, academia and researcher in the field of network modeling and performance analysis, traffic engineering and network characterization, optical communications and networking, network resource management and quality of services (QoS) control, video communications over next generation Internet, wireless/mobile communications and networking, switching and routing algorithm, ad-hoc networks, radio frequency and electro-optical sensor networks, and information assurance and network security. He has published more than 150 research papers in international journals. Dr Zhang is a Senior Member of IEEE and also he is the associate editor of the Journal of Computer Communications and the editor for Journal of Communications, Journal of Networks, Journal of Multimedia and Journal of Ultra Wideband Communications and Systems. Currently, Prof Zhang is working on several R&D projects in the area of Ultra-wideband mobile ad hoc network, Biomedical imaging using ultra wideband, extraction of biometric signatures for forensic identification and multi-layer security architecture in next generation wireless communication networks.